\documentclass{article}
\usepackage{PRIMEarxiv}
\usepackage{amsmath}
\usepackage[utf8]{inputenc} % allow utf-8 input
\usepackage[T1]{fontenc}    % use 8-bit T1 fonts
\usepackage{hyperref}       % hyperlinks
\usepackage{url}            % simple URL typesetting
\usepackage{booktabs}       % professional-quality tables
\usepackage{amsfonts}       % blackboard math symbols
\usepackage{nicefrac}       % compact symbols for 1/2, etc.
\usepackage{microtype}      % microtypography
\usepackage{lipsum}
\usepackage{fancyhdr}       % header
\usepackage{graphicx}       % graphics
\usepackage{xcolor}
\graphicspath{{media/}}     % organize your images and other figures under media/ folder
\usepackage{multicol}
\usepackage{longtable}
\usepackage{chemformula}
\title{Large cities are less efficient for sustainable transport: The ABC  of mobility}

\author{
  Rafael Prieto-Curiel\\
  Complexity Science Hub \\ Josefstaedter Strasse 39 \\ 1080 Vienna, Austria \\
  \texttt{prieto-curiel@csh.ac.at} \\
   \And
  Juan P. Ospina \\
  Research in Spatial Economics \\ Universidad EAFIT \\ Medellín, Antioquia, Colombia\\
  \texttt{jospinaz@eafit.edu.co}\\
}

\begin{document}
\maketitle

\begin{abstract}

The use of cars in cities has many negative impacts on its population, including pollution, noise and the use of space. Yet, detecting factors that reduce automobile dependency is a serious challenge, particularly across different regions. Here we model the use of different modes of transport in a city by aggregating active mobility (A), public transport (B) and cars (C), thus expressing the modal share of a city by its ABC triplet. Data for nearly 800 cities across 60 countries is used to model car use and its relationship with city size and income. Our findings suggest that outside the US, longer distances experienced in large cities reduce the propensity of active mobility and of cars, but public transport is more prominent. For cities in the US, roughly 90\% of its mobility depends on cars, regardless of city size. Further, income is strongly related to automobile dependency. Results show that a city with twice the income has 37\% more journeys by car.
\end{abstract}

\section{Introduction}

%%%% cars are super bad
{
Cars present significant challenges to cities due to their resource requirements, lifestyle implications, and environmental impact. The increase in the motorisation rate is a massive burden on cities, leading to issues such as unsafe roads, noise, pollution, sedentary habits, inefficient use of public space, and costly infrastructure primarily dedicated to parked cars \cite{banister2011cities, rabl2012benefits}. Additionally, transportation-related pollution claims the lives of thousands, and road accidents annually result in the loss of 1.3 million lives, disproportionately affecting pedestrians, cyclists, and motorcyclists  \cite{choma2021health}. 
}

%%%% cars are super bad 2
{
Rapid urbanisation and car-centric infrastructure have resulted in sprawling low-density suburbs, contributing to longer commuting distances and heightened environmental pollution \cite{Marchetti1994, MATTIOLICarUse2020}. Traffic congestion further exacerbates the situation, partly influenced by private cars and other drivers \cite{roughgarden2002bad, YounPoA, RoughgardenPoP, pas1997braess, dafermos1984some}. The pathway towards sustainable mobility is quite challenging in cities where most journeys are on private cars. Despite this, policymakers still invest disproportionally in car infrastructure, inadvertently reinforcing incentives for private vehicle usage \cite{cervero2002built, duranton2011fundamental, brand2021climate, keall2018reductions, schafer2000future, mogridge1997self, prieto2021paradox}. 
}

%%%% yet, active mobility and determinants of people moving
{
Encouraging active mobility and public transport as alternatives to driving presents numerous physical and environmental advantages. However, changing travel behaviour is exceptionally challenging \cite{schafer2000future, shaw2020mode, keall2015increasing, keall2018reductions, pucher2004public, kristal2020we, brand2021climate}. People often choose to drive due to factors such as infrastructure availability, expected travel time, accessibility, comfort, and security \cite{WANG20161, HERRMANNLUNECKE2021192, cervero2002built, de2019satisfying, wild2019cyclists, Kuhnimhof2012}. The choice of transportation mode significantly impacts cities, potentially leading to either greener or more polluted environments. In the US, driving accounts for approximately 75\% of household-level \ch{CO2} emissions, while public transportation, residential heating, and electricity consumption contribute to the remaining 25\% \cite{glaeser2010greenness}. And even with the growth of the electric cars market, the burden associated with motorised mobility will remain, including their manufacture, infrastructure demands, congestion, particle pollution produced by tyre wear, and others \cite{prieto2021paradox}. %Cities could gain massive permeable and biodiverse green spaces by de-paving redundant parking spaces \cite{croeser2022finding}. 
}

%%%% city size and its impact. scaling
{
Promoting sustainable transportation involves prioritising mobility beyond cars, emphasising fewer and shorter trips, non-motorised modes like walking and cycling, and shared transportation through public transit \cite{saeidizand2022revisiting}. Active mobility for short distances and public transport for longer distances are the key answers to sustainable mobility \cite{brand2021climate}. Yet, active mobility and public transport face significant barriers across cities of different sizes. Long commuting distances in medium and large cities make active mobility challenging. On the other hand, public transport requires sufficient passenger volume to provide frequent service, and its availability is influenced by population density and city size \cite{downs1962law, mohring1972optimization, levinson2007planning, bar2013model, newman1989cities, wu2021urban}. Smaller cities often lack public transport options, while medium and large cities discourage active mobility due to long travel distances. City size strongly influences transportation choices and the feasibility of different modes of transport. 
}

%%%% 
{
City size is crucial in various aspects of social life and infrastructure. Larger cities tend to attract more migrants and highly skilled individuals and exhibit higher income, innovation, and crime rates \cite{ScalingMigrationRPC, keuschnigg2019urban, GrowthBettencourt, UrbanScalingWest, louf2014congestion, pumain2004scaling, zhou2020gini}. Regarding mobility, larger cities experience more road accidents, although they are less likely to be fatal \cite{cabrera2020uncovering}. However, determining how city size affects modal share is still an open question. Describing the ideal mobility for a city of a given size is not trivial. People tend to switch to faster commuting methods which could be a private vehicle in most cases \cite{prieto2022ubiquitous}. 
}

%%%% data. Juanpa
{
We compiled data for 797 cities across 62 countries to quantify modal share across different cities. We used several sources, including the European Platform on Mobility Management (EPOMM), Local Governments for Sustainability (ICLEI), Deloitte City Mobility Index, World Conference on Transport Research, and the US Census, among others \cite{epomm, ICLEI, ABSCensus2016, Pareekh2017} (SM A). Most data used is before the covid pandemic to avoid possible mobility shocks due to the lockdown. 
}

%%%% assumptions
{
We analyse the impact of city size on transportation patterns, focusing on comparisons within and outside the United States (US). While data limitations and a bias toward European cities exist, our dataset includes nearly 100 observations from non-US and non-European cities. This provides insights into modal shares and their influence on transport-related emissions globally. We find similarities in modal shares across regions, such as Canadian cities displaying similar car dependency to their US counterparts. Many Latin American cities show a more balanced distribution between active mobility, public transport, and driving. Active mobility is common in small cities outside the US, while public transport is preferred in larger cities. In contrast, car journeys dominate the US, regardless of city size. Additionally, we highlight the direct impact of income on modal shares, with higher income levels correlating with increased car usage and reduced reliance on public transport. 
}

\section{Results}

%%%% the ABC of mobility
{
To simplify the analysis of transportation modes in various cities, we use the ABC scheme, which categorises them as Active Mobility (A), Bus or Public Transport (B), and Cars (C). For city $i$, we combine the proportions of active mobility (including walking, cycling, etc.) into category $A_i$, public transport journeys (including Metro, BRT, etc.) into category $B_i$, and private motorised journeys (including cars, taxis, etc.) into category $C_i$. These categories are represented on a ternary plot (Figure \ref{Ternary}), such that $A_i + B_i + C_i = 100\%$  \cite{CiteTernaryR, CiteR}. This representation condenses diverse transport modes into a triangle, with each corner representing a scenario where everyone uses a single mode of transport \cite{szell2018crowdsourced, prieto2022ubiquitous}. 

\begin{figure}
\begin{center}
\includegraphics[width = 0.6\linewidth]{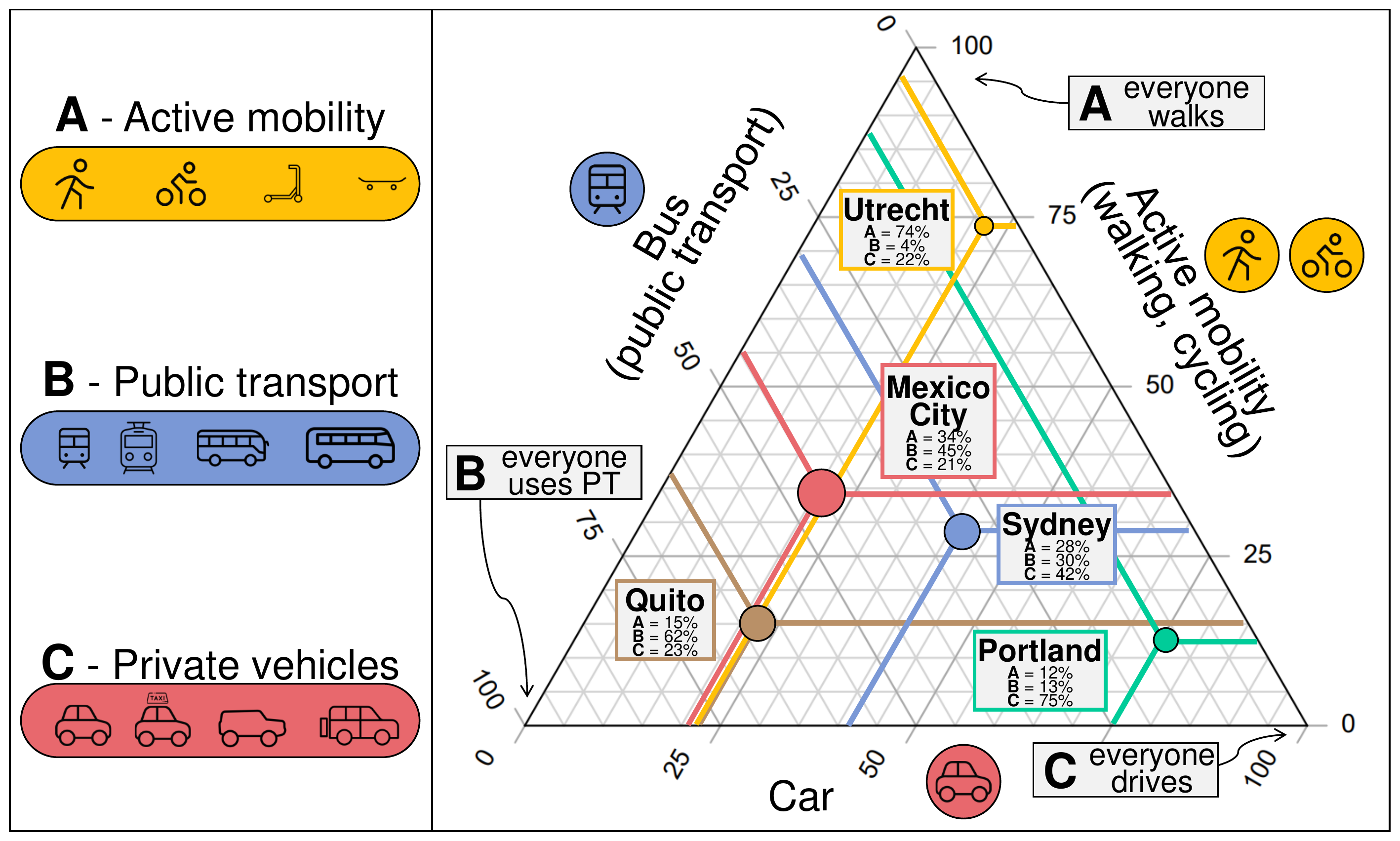}
\end{center}
\caption{We aggregate different modes of transport into a category of Active mobility, public transport and private vehicles. Then, the share of journeys to work that are active mobility $A_i$ (walking, cycling, skating and others) in the right of the triangle, on public transport $B_i$ (bus, BRT, Metro, Tram and other shared systems) on the left, and by private cars $C_i$ (including taxis and mobility apps) on the bottom. Each dot in the ABC triangle represents the modal share of a city. For each city, the level of active mobility is observed on the right, buses on the left, and cars on the bottom.} \label{Ternary}
\end{figure}
}

%%% some description of the mobility numbers with figure
{
Comparing modal share across cities is challenging since it depends on the definition of the urban area, the type of journeys, the different modes of transport, and how multi-modal journeys are reported \cite{saeidizand2022revisiting}. Our analysis compares the modal share using data from 797 cities across 62 countries. We find that 70\% of the journeys are by car, 19\% are walking or cycling (active mobility), and 11\% are by public transport. However, there are substantial differences between regions (Figure \ref{MsFigure2}). In USA and Canada, 95\% of the journeys are by car, and only 3\% are active mobility. In Europe, active mobility is not preferred since more than half of its trips are by car. In some metropolitan areas like Greater Manchester, we find that 71\% of their journeys are on a private vehicle. Conversely, in African cities, nearly half of the trips are active mobility.

\begin{figure}
\begin{center}
\includegraphics[width = 0.6\linewidth]{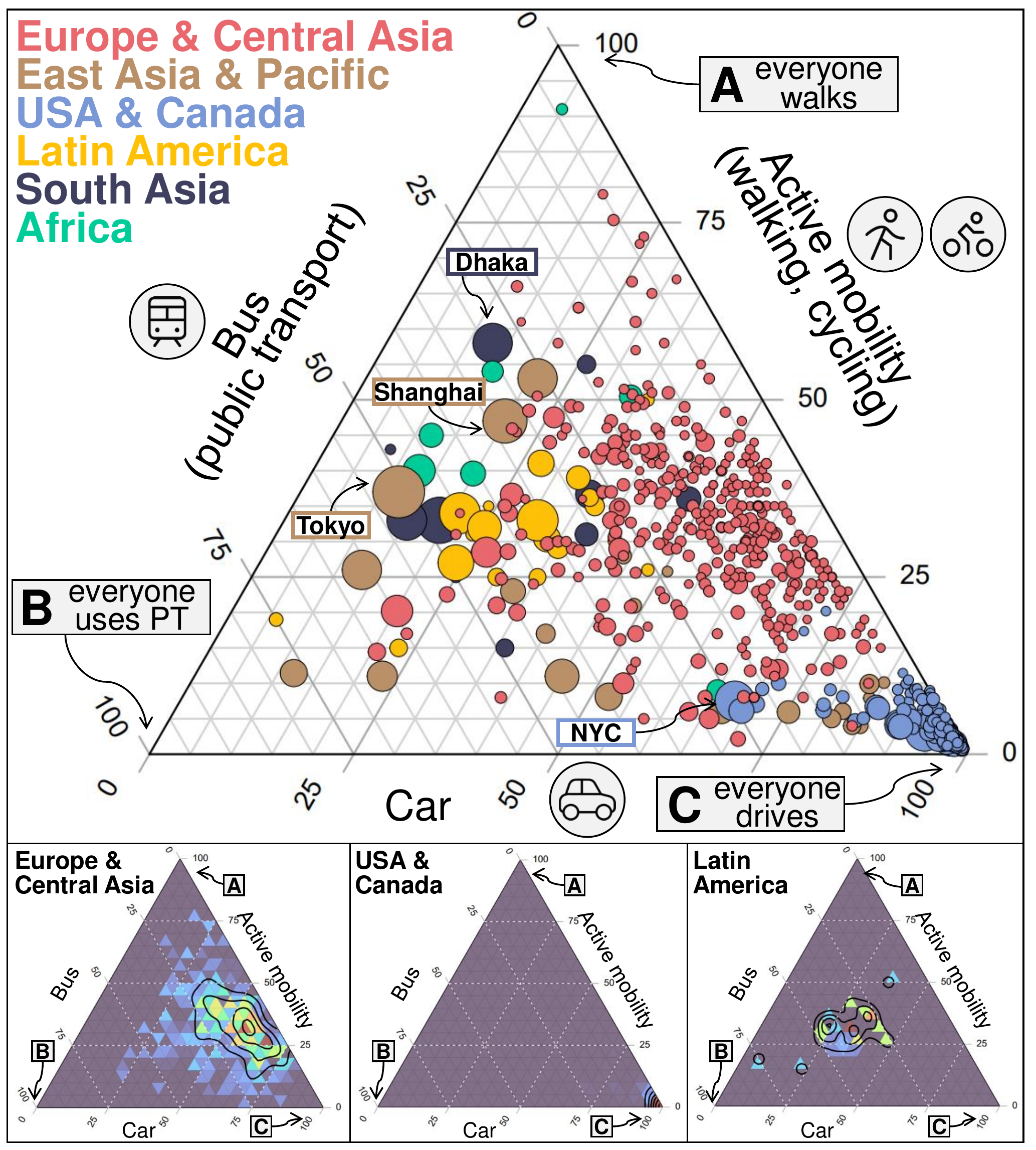}
\end{center}
\caption{ABC modal share for 797 cities (top) where the size of the disc is proportional to the city's population, and the colour corresponds to the region. Density and contours of cities observed (bottom) in Europe and Central Asia, USA and Canada and in Latin America.} \label{MsFigure2}
\end{figure}
}

%%% scaling results with scaling figure
{
Outside the US, city size has a significant influence on transportation patterns. Approximately one-third of all journeys are attributed to active mobility, such as walking or cycling. However, in cities with populations exceeding one million, longer distances discourage people from choosing these modes of transport. Active mobility and car journeys are common in small cities, while public transport options are limited. However, as cities grow, there is a notable shift towards public transport as the primary mode of transportation (Figure \ref{ScalingFigure}). For cities with fewer than 100,000 inhabitants, public transport accounts for approximately 10\% of journeys. This percentage increases to around 25\% for cities with around one million people and surpasses 40\% for cities with over 20 million inhabitants. 

\begin{figure}
\begin{center}
\includegraphics[width = 0.8\linewidth]{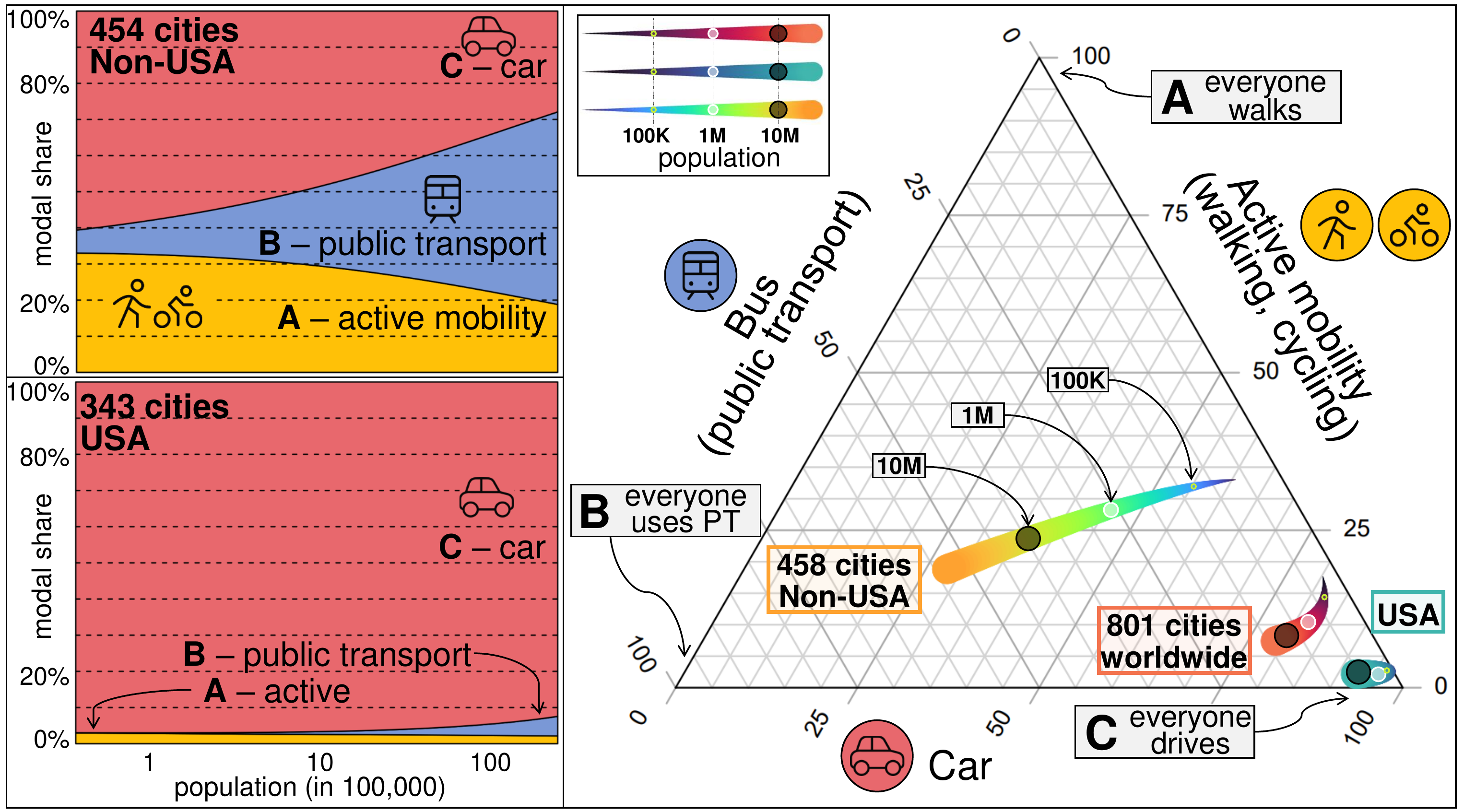}
\end{center}
\caption{Left - Impact of city size on modal share for cities outside the US (top) and inside the US (bottom). Right - Impact of city size across the ABC model for cities between 20,000 and 40 million inhabitants.} \label{ScalingFigure}
\end{figure}
}

%%%% difference US vs the rest
{
In contrast, the US exhibits minimal variation in modal share across cities of different sizes. Regardless of city size, approximately 95\% of mobility relies on private vehicles. There is a slight trade-off in the remaining 5\%, with a slightly higher proportion of active mobility in small cities and public transport in larger cities. The majority of cities in the US have been designed with a strong reliance on cars for transportation. While cities like New York City and Boulder have developed alternative mobility options, most cities in the US heavily depend on vehicles. Surprisingly, statistics show that regardless of city size, individuals in the US have an equal likelihood of driving. Approximately 90\% of the population drives, irrespective of whether they reside in a small or large city. The remaining 10\% predominantly consists of active mobility for smaller cities and public transport for larger cities.
}

%%%% mention walking VS cycling
{
Walking as a mode of transport works better in small cities, but as distances increase, walking is less frequent. For cities outside the US, we estimate that when it doubles its size, the probability that a person walks as a mode of transport is reduced by 5\%. Something similar happens with cycling, although it is even more sensitive to large distances. When a city doubles its size, the probability that a person cycles drops 11\% (see SM B). For cities in the US, walking and cycling are very limited, but they decrease as cities double in size by 6\% and 3\%, respectively. 
}

%%%% income
{
High income is correlated with car dependency (Figure \ref{IncomeFigure}). As income rises, the frequency of active mobility and public transport decreases. Our analysis reveals a relationship described as $C_i \propto Y_i^{\gamma_C}$, where $C_i$ represents car journeys and $Y_i$ represents the income of a city. The exponent $\gamma_C$ is estimated to be $0.455 \pm 0.018$, which implies that a city with twice the income (holding other factors constant) is associated with a 37\% increase in car journeys, primarily reducing the number of public transport journeys (SM C).

\begin{figure}
\begin{center}
\includegraphics[width = 0.8\linewidth]{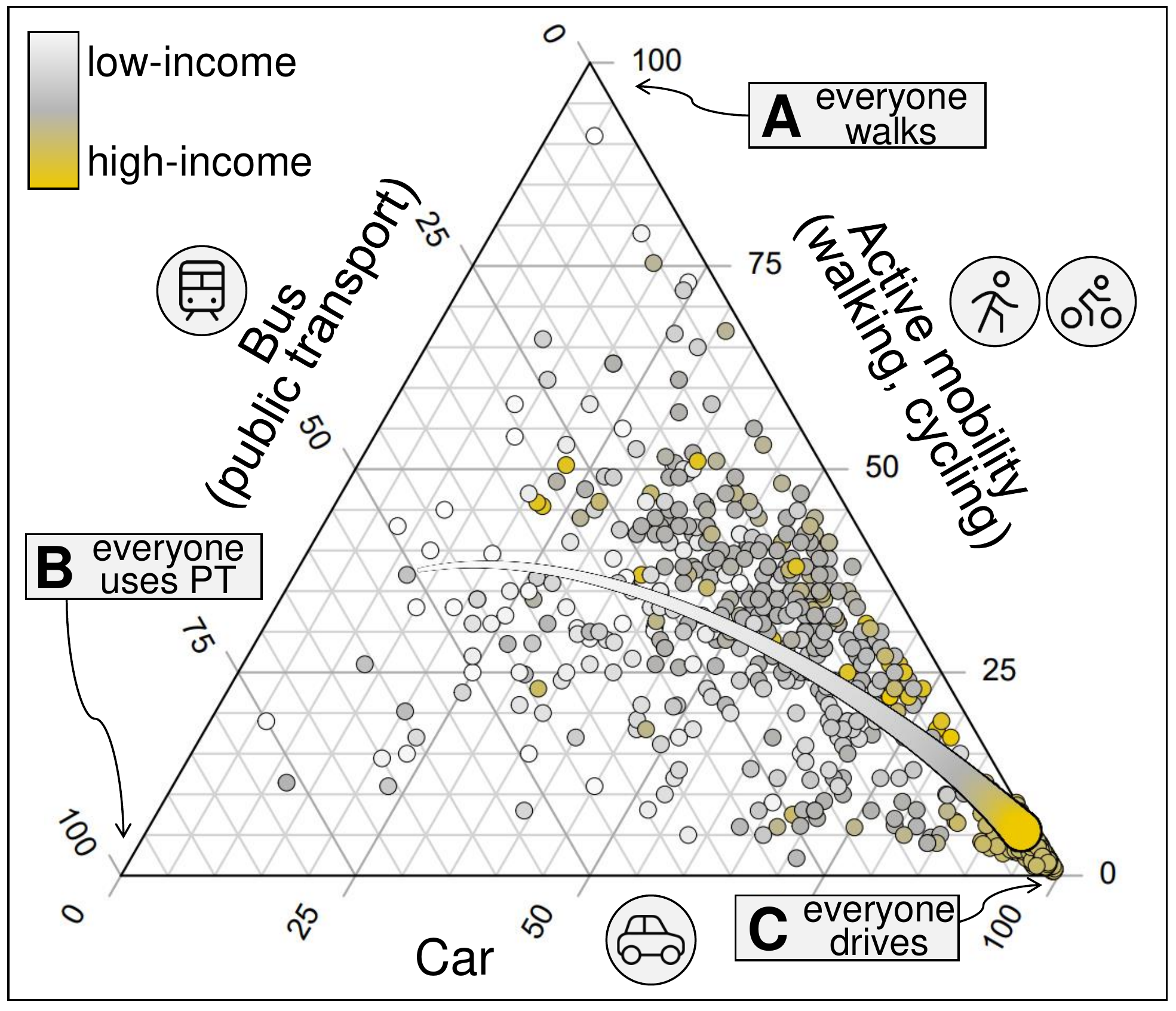}
\end{center}
\caption{Impact of income on the modal share distribution of a city. Each disc represents a city, and its colour represents the income per person at the country level in 2021. The curve represents the predicted modal share of cities with varying incomes. } \label{IncomeFigure}
\end{figure}

}

\section{Discussion}

%%% intro p
{
By grouping various forms of active mobility, public transport, and private cars into three overarching categories, we can effectively capture the modal share of cities using the ABC scheme. This representation allows us to observe distinct variations in mobility patterns across different cities and countries. While certain cities, notably in the US, heavily depend on private cars, alternative modes of transportation are more widely embraced in other regions.
}

%%%% challenges wrt data
{
Other studies comparing the modal share across cities rely on less than 50 metropolitan areas \cite{saeidizand2022revisiting, MobilityCities2015, CityTransit}. Our novel data set includes information on the modal share of 797 cities worldwide. However, this data needs to be more representative regarding geographical distribution or city size. Most observations are from Europe and the USA, while data from middle and low-income countries is limited. Obtaining data on modal share is challenging, as it is only available for a subset of cities. Additionally, comparability among the data poses a challenge, as some surveys capture specific types of motorised mobility (such as motorbikes, cycling, and scooters). In contrast, others provide aggregate categories, which makes it difficult to separate and analyse other attributes of urban mobility. 
}

%%%% scaling and active mobility and PT
{
City size plays a crucial role in shaping transportation patterns and choices. However, it is worth noting that mobility patterns are also affected by the design and layout of cities. In larger cities, active mobility options (walking or cycling) can be less attractive due to longer distances, yet, driving also faces increased costs. Our results show that as a city's population doubles in non-US cities, the likelihood of individuals choosing active mobility decreases by 10\%, and the probability of using a private car decreases by 7\%. However, public transport compensates for these challenges. As cities grow, the probability of using public transport increases by 18\%. Across cities of varying sizes in the US, there is minimal variation in the modal share of transportation. Approximately 95\% of mobility in the US is dominated by private vehicle usage, regardless of the city's size. The remaining 5\% represents a minor trade-off, with small cities showing a slightly higher proportion of active mobility and larger cities relying more on public transport.

Income level also influences transportation choices. Additionally, when a city's income doubles, more car journeys are expected, primarily at the expense of public transport usage. However, compelling examples around the globe illustrate how income and reliance on automobiles can be decoupled. In cities such as Zurich, Copenhagen, Tokyo, and Amsterdam, diverse transportation options, including cycling paths, tramways, underground railways, buses, and pedestrian pathways, intertwine with the urban landscape. These cities serve as shining models where the wealthiest and most economically vibrant urban centres are at the forefront of reshaping transportation infrastructure. Their objective is to foster neighbourhoods prioritising safety, well-being, and a more enjoyable living environment for their inhabitants.
}

%%%% for outisde the US
{
The current balance between cars, public transport, and active mobility in certain cities does not guarantee a sustainable future, nor does it imply that cities are progressing in the right direction. For instance, in cities such as Mexico City or Bogota, the public transport system is overwhelmed, uncomfortable, and lacks security, leading to an estimated one in five individuals opting for private cars if they could afford them \cite{prieto2021paradox, katz2010levels, varghese2016measuring, samson2017crowd, lastra2015mi, Guzman2018}. The primary cause of having a small car dependency in many cities is income rather than the availability of favourable conditions for sustainable mobility. Many public transport users have no alternative options for travel, making them captive riders \cite{bar2013model, Bocarejo2012}. Therefore, achieving sustainable mobility involves promoting more active mobility and public transport in car-dominated cities while reducing the attractiveness of driving as a mode of transportation \cite{saeidizand2022revisiting}.
}

%%% No replacement of traditional methods
{
%It is important to note that our intention is not to dismiss or replace the knowledge and insights provided by traditional transportation experts.
Our findings reinforce that our cities must adequately address their transportation-related challenges. As the repercussions of climate change continue to inflict harm upon our planet and its inhabitants, urban environments centred around oversized, single-occupant vehicles appear increasingly out of sync with our evolving needs. It is increasingly evident that a fundamental shift in approach is required. Current strategies and practices must overcome these urban challenges. By recognising the need for a paradigm shift, we can embark on a transformative journey to develop innovative solutions and embrace new approaches to transportation that can effectively tackle these pressing issues.}

%%% Not a war against cars.
{
%Neither the aim is not to eliminate cars but rather to create a balanced approach where cars coexist with alternative modes of transportation, such as public transit, walking, and cycling. 
By promoting a diverse transportation ecosystem, cities can improve mobility options, reduce congestion, enhance accessibility, and create more vibrant and livable urban environments for their residents. An ever-growing body of evidence highlights the potential environmental and public health benefits of reducing car dependency. This shift encompasses a reduction in pollution and an increase in active modes of transportation, such as walking, biking, and other physical activities. By embracing alternative means of transportation, we can foster a healthier and more sustainable future for ourselves and our planet.
}

\section{Methods}

%the ABC model
{
In the ABC scheme, we assign individuals' city trips to Active, Bus, or Car categories. This allocation is based on the assumption that everyone travels and can be categorised accordingly. Let $P_i$ be the population of the city $i$, and let $A_i$ be the proportion of active mobility journeys, including walking, cycling, and others, $B_i$ the proportion of Bus journeys, including other types of public transport, such as Tram, Metro and different public transport journeys and let $C_i$ be the proportion of car journeys in the city $i$ including private car drivers, taxis and others. Multimodal journeys are assigned first to the car and then to public transport categories if either is used for some part of the journey. We model the weekly travel intensity, $\mu$, so that a person expects to make $\mu$ journeys weekly. The modal share distribution is determined by the probabilities corresponding to each city's modal share. Hence, the expected number of car journeys for a person in city $i$ is $\mu P_i  C_i$, and the same applies to active mobility and public transport. 

We investigate the influence of city size on the modal share distribution by expressing the frequency of active mobility journeys as $\mu P_i A_i \propto P_i^{\beta_A}$, for bus journeys as $\mu P_i B_i \propto P_i^{\beta_B}$ and for car journeys as $\mu P_i C_i \propto P_i^{\beta_C}$.  
This implies that the probability of an individual in the city $i$ choosing active mobility is proportional to $A_i \propto P_i^{\beta_A-1}$, and this probability may vary across different city sizes. If we obtain that $\beta_A = 1$, then the probability of selecting active mobility would be the same across cities of various sizes. With $\beta_A > 1$, active mobility would be more prevalent in larger cities, while $\beta_A < 1$ indicates a lower frequency of active mobility in large cities. Although conventional mobility surveys only consider people older than five years, we consider the total number of people living in a city $i$ for our estimations.
}

%%%% Result Number
{
Data for 797 metropolitan areas is used to calibrate the parameters. For some cities, we have more than one year of data available, up to seven years for Graz and six for Vienna. We use the latest observation for the 797 cities to estimate the values of the parameters (Table \ref{Parameters}). We then subdivide the results by comparing 454 cities outside the US and 343 cities in the US. Comparing two cities $C_1$ and $C_2$, where $P_1 = 2 P_2$, so that city $C_1$ has twice the population of city $C_2$, we get that the frequency of car journeys decreases 5\% in the large city, and also active mobility is reduced by 2\%, but they compensate with an increase in 18\% in the frequency of public transport journeys. Thus, the probability that a person uses public transport increases by 18\% for a city with twice the population.
}

%%%% results 2022/12/19 with all cities
{
\begin{table}
\centering
 \begin{tabular}{|r |c c|cc|cc|} 
 \hline
  & \multicolumn{2}{c|}{797 cities} & \multicolumn{2}{c|}{454 non-US cities} & \multicolumn{2}{c|}{343 US cities} \\
 Mode & $\alpha$ &  $\beta$& $\alpha$ &  $\beta$& $\alpha$ &  $\beta$ \\
 \hline
$A$ - active & $0.6389$   & $0.8528$   & $0.6666$ & $0.9289$   &  $0.0408$ & $0.9646$  \\
             & $(0.3188)$ & $(0.0319)$ & $(0.3327)$ & $(0.0153)$ & $(0.0203)$& $(0.0307)$  \\
\hline             
$B$ - bus & $0.0039$  & $1.1737$  & $0.0071$ & $1.2437$  &  $1.4e-5$ & $1.4645$ \\
          & $(0.0019)$& $(0.0483)$& $(0.0035)$ & $(0.0167)$& $(7.1e-6)$ & $(0.0669)$\\
\hline
$C$ - car & $1.1444$ & $0.9532$     & $1.5371$ & $0.9043$  &  $1.0203$  &$0.9949$  \\ 
          & $(0.5711)$ & $(0.0118)$ & $(0.7671)$ & $(0.0113)$& $(0.5092)$& $(0.0018)$ \\
 \hline
 \end{tabular} 
 \caption{Parameters estimated for the number of $A$ Active journeys, $B$ bus journeys, and $C$ car journeys.} \label{Parameters}
\end{table}
}

\subsection{The ABC data and its comparability}  

%%%% how the data
{
Modal share data is typically gathered through travel surveys, which primarily focus on the primary mode of transportation a person employs for each trip during a typical weekday. These surveys may differentiate between car drivers, car passengers (including taxi passengers), and motorbike riders. However, in some cases, surveys do not provide distinctions between different types of private vehicles. To ensure comparability, we combine all private motorised trips into a category referred to as ``Car" journeys. Similarly, public transport surveys may encompass various modes such as metro, tram, bus, and other forms of public transportation. To facilitate uniformity, we aggregate all types of public transport into a single category known as ``Bus". Active mobility encompasses a range of physical activities, including walking, cycling, electric bicycles, skating, rollerblading, and more. To consolidate the various forms of active mobility, we combine them into a category called ``Active mobility". By reducing transportation modes into these broad categories, we create a standardised framework for analysing and comparing modal share data across cities and regions. 
}

%%% metro areas
{
The definition of cities or metropolitan areas is crucial in determining many urban indicators and also the mobility types observed across different locations \cite{arcaute2015constructing}. Surveys that rely on spatial or political boundaries to define cities can introduce spatial biases into the data. To mitigate this, we focus on urban areas and their respective populations as the primary observation units. This approach allows us to exclude observations that do not precisely align with a specific metropolitan area, ensuring greater accuracy and consistency in our analysis.
}

%%% not alligned
{
The available data for analysis is not consistently aligned in terms of time. Travel surveys are conducted only occasionally for most cities, which presents a challenge regarding temporal consistency. To address this, we employ the most recent data for each city in our analysis. To avoid any disruptions caused by the COVID-19 lockdowns, we specifically consider data from 2019 or earlier for US cities. Overall, our compiled dataset includes information from 797 cities. In 76\% of these cities, the data is from 2010 or more recent, and in 93\% of the cities, the data is from 2005 or more recent. Despite the temporal misalignment, we assume that changes in the modal share at the city level occur gradually, allowing for meaningful analysis. Consequently, variations observed in the modal share across cities are attributed to underlying structural and social differences rather than temporal or other survey-specific discrepancies.
}

%%% sources
{
The modal split analysis relies on data primarily sourced from reputable organisations such as the European Platform on Mobility Management (EPOMM), Deloitte City Mobility Index, World Conference on Transport Research, US Census, EcoMobility Alliance Cities, TUMI's Global Mobility Challenge 2018 Winners, EcoLogistics project cities, and CitiesSHIFT project cities. However, the data quality may vary due to changes in definitions and data collection methods across different cities. In some cases, actual data may be available, while in others, estimates based on alternative methodologies may be utilised.
}

%%%% triplet
{
To represent the modal share of a city, we assign fractions to different types of journeys: active mobility ($A_i$), bus journeys ($B_i$), and car journeys ($C_i$). These fractions are constrained to the range $[0,1]$, ensuring that their sum equals 100\%. This data structure allows us to visualise the modal share of a city as a triplet $(A_i, B_i, C_i)$. We display the modal share on a two-dimensional triangle known as a 2-simplex by utilising ternary data representation \cite{CiteTernaryR, CiteR}. It provides a visual framework to understand the relative proportions of active mobility, bus journeys, and car journeys within a city.
}

\subsection{Parameter estimation}

%%%% scaling
{
In the method to estimate the scaling relationship between city size and the modal share of a city, we assume that each person takes $\mu $ journeys per week and that the probability of choosing some transport mode corresponds to the city's modal share. Thus, we estimate that on a week, a city with $P_i$ individuals expects $\mu P_i A_i$ active mobility journeys, $\mu P_i B_i$ public transport trips, and $\mu P_i C_i$ car journeys. These values correspond to the city's expected weekly $A$, $B$, and $C$ journeys. Then, we express the number of active mobility journeys as $\alpha_A P_i^{\beta_A}$ for some values of $\alpha_A$ and $\beta_A$ and similarly for public transport and car journeys. The values and standard error of $\alpha_A$ and $\beta_A$ are estimated through a regression (Figure \ref{Scaling}).

\begin{figure}
\begin{center}
\includegraphics[width = 0.5\linewidth]{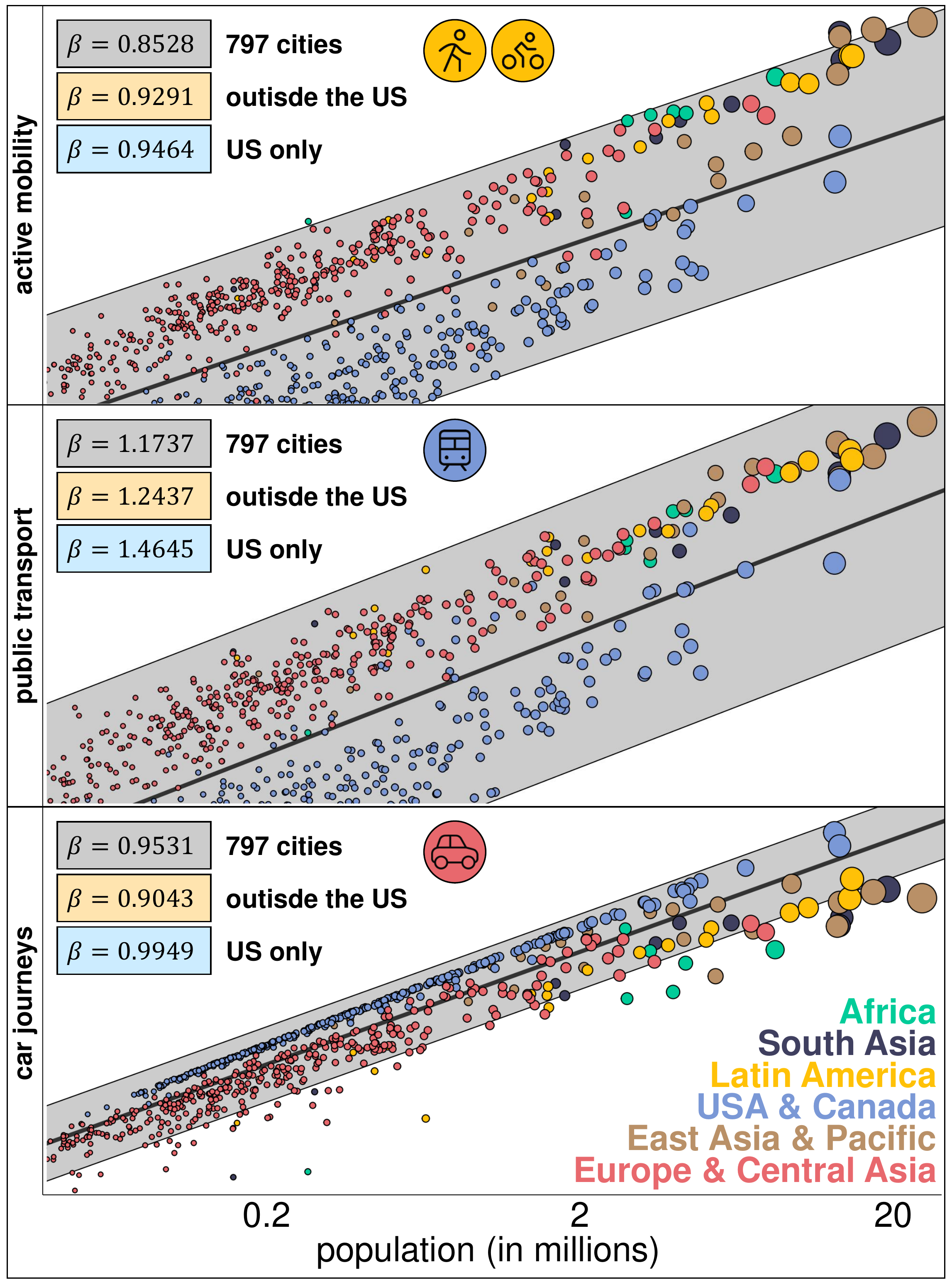}
\end{center}
\caption{Public transport journeys (vertical axis) vs population (horizontal axis). The results of the coefficients are obtained for all cities (model displayed) and by considering only cities outside and inside the US.} \label{Scaling}
\end{figure}
}

%%% income
{
To estimate the impact of income, we consider the GDP per person, compiled by the World Bank for 2021. We then fit the regression $\log C_i = \alpha + \beta \log Y_i$, where $Y_i$ corresponds to the income per person in that country (expressed in current US dollars). Based on that equation, we obtain that the car share in city $i$ can be expressed as
\begin{equation}
C_i = 0.005 Y_i^{0.455},
\end{equation}
meaning that with a higher income, more car journeys are expected. Similarly, we express the number of active mobility journeys as $A_i \propto Y_i^{-0.847}$, and the number of public transport journeys as $B_i \propto Y_i^{-1.435}$, meaning that with more income, fewer active mobility and public transport journeys are expected.
}

\section{Supplementary Materials}

\subsection*{A - List of sources} %%%% Supplementary A

\tiny 
\begin{longtable}{|l|l|l|} 
\caption{\small List of sources, number of cities included on each source, and references.}\label{tab:sources}\\
    \hline
    \multicolumn{1}{|c|}{\textbf{Source}} & \multicolumn{1}{c|}{\textbf{\#\_cities}} & \multicolumn{1}{c|}{\textbf{Reference}} \\
    \endfirsthead
    \caption* {Table \ref{tab:sources} (Continued): List of sources.}\\ \hline
    \multicolumn{1}{|c|}{\textbf{Source}} & \multicolumn{1}{c|}{\textbf{N\_cities}} & \multicolumn{1}{c|}{\textbf{Reference}} \\ \hline 
    \endhead
    \hline
    2011 Transport characteristics Survey & 1 & \cite{HongKong2011} \\
ABS census data 2016 journey to work & 6 & \cite{ABSCensus2016} \\
Background to the Transport Evidence informing the Ipswich Local Plan & 2 & \cite{Ipswich2015} \\
Beijing Yearbook 2011 & 1 & \cite{Beijing2011} \\
Bonn - Modal split & 1 & \cite{Bonn2016} \\
Census of India 2011 & 4 & \cite{India2011} \\
Census Profile Canada & 11 & \cite{Canada2016} \\
Central Bureau of Statistics - The Netherlands & 1 & \cite{Amersfoort2019} \\
Cerema & 1 & \cite{Groningen2008} \\
Change trends in the use of passenger cars on urban trips: car-pooling in Gdynia & 1 & \cite{Hebel2017} \\
Civitas & 9 & \cite{Graz2020} \\
CODATU & 1 & \cite{Ferrara2011} \\
Cycling Academics. Bringing science on cycling to practice and back & 1 & \cite{Brommelstroet2013} \\
Darmstadt - Modal split & 1 & \cite{Darmstadt2016} \\
Dataplatform - The Netherlands & 3 & \cite{dataplatform2011} \\
Deloitte City Mobility Index & 4 & \cite{Helsinki2018} \\
Dhaka mobility study & 1 & \cite{Ali2021} \\
Edinburgh City Mobility Plan-2030 & 1 & \cite{Edinburgh2021} \\
ELTIS - The Urban Mobility Observatory & 4 & \cite{Basel2015} \\
EOD Bogotá & 2 & \cite{BogotaEOD} \\
Erfurt.de – das offizielle Stadtportal & 3 & \cite{Erfurt2013} \\
Ergebnis der Verkehrserhebung 2012 & 2 & \cite{Linz2014} \\
Fürth - Modal split & 1 & \cite{Furth2016} \\
Global BRT data & 4 & \cite{Enschede2008} \\
GPSM\_Discover Leipzig by Sustainable Transport & 2 & \cite{Leipzig2016} \\
Grundsatzbeschluss zur Förderung des Radverkehrs in der Stadt Fürth & 1 & \cite{Furth2018} \\
Haushaltsbefragung zum Mobilitätsverhalten der Dortmunder Bevölkerung 2013 & 1 & \cite{Dortmund2014} \\
Haushaltsbefragung zum Mobilitätsverhalten in Essen 2019 & 2 & \cite{Essen2019} \\
How did Bicycle Share Increase in Vitoria-Gasteiz & 3 & \cite{Monzon2016} \\
Korea National Transportation DB Report 2013 & 1 & \cite{Korea2013} \\
Leeds Transport Strategy & 1 & \cite{Leeds2021} \\
Liverpool City Region Household Travel Survey & 3 & \cite{Liverpool2018} \\
Local Governments for Sustainability - ICLEI & 33 & \cite{ICLEI} \\
Mobilität in Deutschland MiD & 1 & \cite{Aachen2019} \\
Mobilität in Deutschland 2017 & 2 & \cite{Zadel2018} \\
Mobilität in Nürnberg & 1 & \cite{Nurnberg2004} \\
Mobilitätsbefragung 2020 Stadt Wuppertal & 2 & \cite{Wuppertal2020} \\
Mobilitätsbefragung: Verkehrswende nimmt Fahrt auf & 2 & \cite{Duesseldorf2018} \\
Mobilitätserhebung Vorarlberg 2017 & 3 & \cite{Vorarlberg2017} \\
MOBILITEITSVISIE ALMERE 2020-2030 & 1 & \cite{Almere2020} \\
Mobility Minsk & 1 & \cite{Minsk2016} \\
Mobility survey AMVA 2012 & 1 & \cite{AMVA2012} \\
Mobility survey AMVA 2017 & 1 & \cite{AMVA2018} \\
Mobility survey Manizales 2017 & 1 & \cite{Manizales2017} \\
Mobility survey Montevideo & 1 & \cite{Montevideo2016} \\
Mobility survey Santiago & 1 & \cite{Santiago2012} \\
Mobility survey São Paulo & 1 & \cite{SaoPaulo2019} \\
Mobility survey Valle de Mexico & 1 & \cite{Mexico2017} \\
Monografia UFRJ & 1 & \cite{Pumar2018} \\
Newcastle \- Empower & 1 & \cite{Newcastle2010} \\
Nürnberg \- Modal split & 1 & \cite{Nurnberg2016} \\
NZ Census 2018 & 3 & \cite{NZ2018} \\
Oberhausen - Mobilitätskarte & 1 & \cite{Oberhausen2020} \\
Observatorio de Movilidad de Bogotá & 1 & \cite{BogotaEOD} \\
Offene Daten Konstanz & 3 & \cite{Konstanz2018} \\
Osaka City Hall & 1 & \cite{Osaka2010} \\
Passenger Transport Mode Shares in World Cities & 1 & \cite{Journeys2011} \\
Past, Present and Future Mobility challenges and opportunities in Bucharest & 1 & \cite{Cavoli2017} \\
Piling up or Packaging Policies? An Ex-Post Analysis of Modal Shift in Four Cities & 1 & \cite{Dijk2018} \\
Plan de Déplacements Urbains d’Annemasse Agglo 2014 & 1 & \cite{AnnemasseAgglo2014} \\
PLAN DE DÉPLACEMENTS URBAINS. AMIENS MÉTROPOLE. 2013 - 2023 & 1 & \cite{Amiens2013} \\
POLIS - Cities and regions for transport innovation & 1 & \cite{Eindhoven2008} \\
Quantifying dimensions of Transportation Diversity: A city-based comparative approach & 5 & \cite{Pareekh2017} \\
Realities of Modal Choice in Kuala Lumpur: Transport Planning for the Disadvantaged & 1 & \cite{Mohamad2017} \\
Regeneration strategies and transport improvement in a deprived area & 1 & \cite{Mahieux2017} \\
Regional Travel Survey of Jyväskylä 2009 & 1 & \cite{Jyvaskyla2009} \\
San Sebastián-Donostia. Calidad urbana: El espacio recuperado & 1 & \cite{Sansebas2010} \\
Seoul & 1 & \cite{Hwang2017} \\
Shanghai Yearbook 2009 & 1 & \cite{Shanghai2009} \\
Sheffield transport data & 3 & \cite{Shefield2020} \\
Städtische Mobilität im Vergleich & 2 & \cite{Bern2015} \\
Taipei Yearbook 2010, Taipei City Government & 1 & \cite{Taipei2010} \\
The effect of Public Transport Investment on car ownership & 4 & \cite{Hass-Klau2014} \\
The European Platform on Mobility Management & 405 & \cite{epomm} \\
The Metropolitan Mobility Observatory of Spain & 2 & \cite{Monzon2011} \\
The Potential of cycling for Sustainable Mobility in Metropolitan Regions  & 2 & \cite{Pospischil2014} \\
Tokyo Statistical Yearbook 2009, Japan & 1 & \cite{Tokyo2009} \\
Traffic behavior in Augsburg & 1 & \cite{Augsburg2018} \\
Transforming Travel in Watford: the strategy for 2021-2041 & 1 & \cite{Watford2021} \\
Travel Survey 2011, Land Transport Authority, Singapore & 1 & \cite{Singapore2013} \\
Urban Audit & 63 & \cite{UE_urbanaudit} \\
Urban structure and sustainable modes’ & 1 & \cite{Tennoy2022} \\
US CENSUS DATA - Commuting & 343 & \cite{us_censusreporter} \\
User Perceptions and Attitudes on Sustainable Urban Transport among Young Adults & 1 & \cite{Puhe2014} \\
Utrecht Mobiliteitsplan 2040 & 2 & \cite{Utrecht2021} \\
Wiesbadener Stadtanalysen & 1 & \cite{Wiesbaden2017} \\
World Bank - Sofia-Bulgaria & 1 & \cite{Sofia2010} \\ \hline

\end{longtable}

\subsection*{B - Walking and cycling in cities}

\begin{figure}
\begin{center}
\includegraphics[width = 0.6\linewidth]{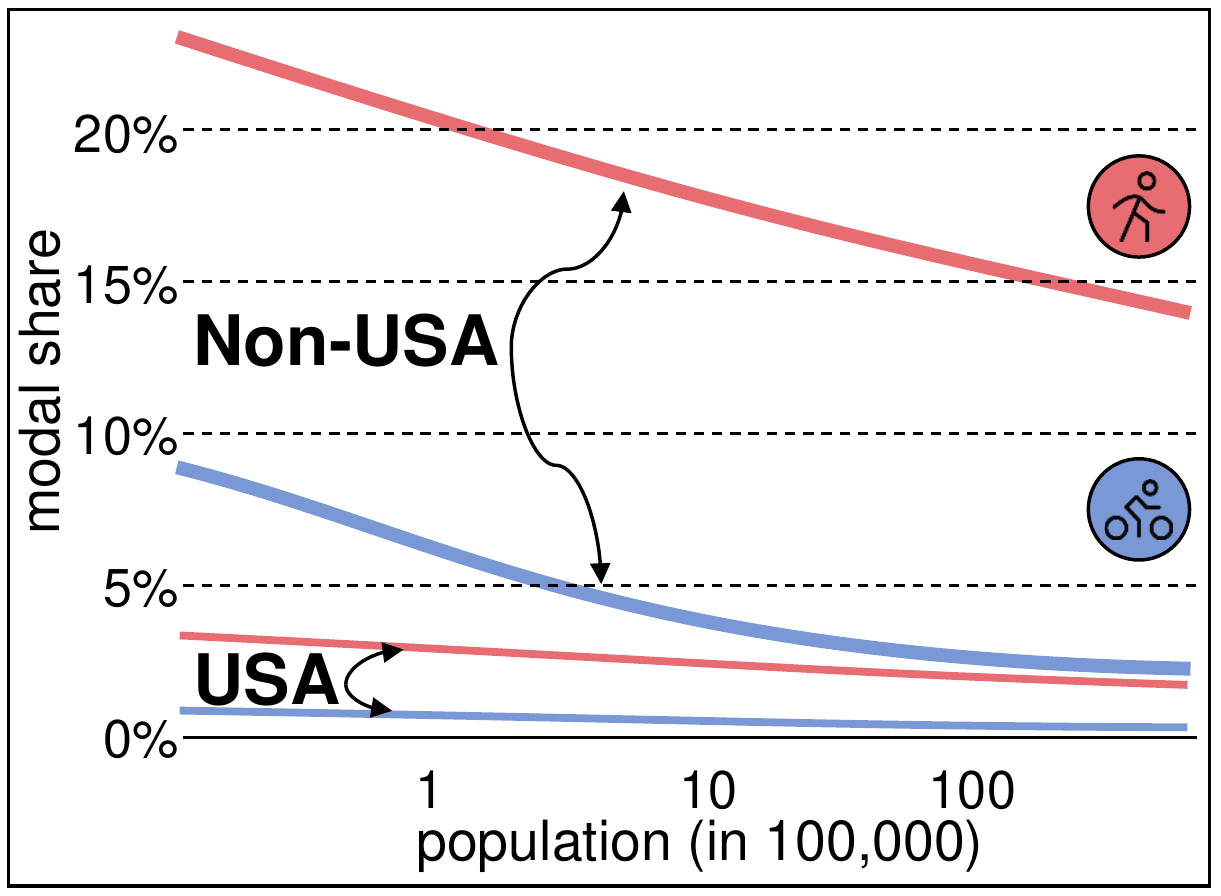}
\end{center}
\caption{Impact of city size (horizontal axis) in the probability of selecting active mobility as a mode of transport (vertical axis). The red line corresponds to the modelled probability that a person walks, and the blue line corresponds to the probability that a person cycles. Thick lines correspond to the model outside the US, and thin lines to the model within US cities. } \label{MsFigure6}
\end{figure}

For 712 out of the 797 cities in the database, we can distinguish the percentage of journeys that are walking and the percentage that are cycling. For the remaining 85 cities, walking, cycling, and other active mobility modes are reported as a single mode, so it is impossible to distinguish between them, and they are dropped from this analysis. Here we explore the scaling impact of city size on the probability that a person decides to walk or that a person chooses to cycle. We divide between cities outside the US and cities within the US for the analysis (Figure \ref{MsFigure6}). 

A person's probability of walking or cycling in a city is lower in bigger cities. However, for cities outside the US, we observe that cycling is more sensitive to city size, so the probability that a person travels by bicycle is smaller in larger cities. This is observed since $\beta_Y = 0.8349 < 0.9409 = \beta_W$. Further, comparing cities of similar size, we observe that it is seven to eight times more likely that a person decides to walk if they are outside the US than if they are within the US. Similarly, a person is between eight and 14 times more likely to use a bicycle if the person is outside the US than if the person is within the US. 

\begin{table}
\centering
 \begin{tabular}{|r |cc|cc|} 
 \hline
  &  \multicolumn{2}{c|}{non-US cities} & \multicolumn{2}{c|}{US cities} \\
 Mode & $\alpha$ &  $\beta$& $\alpha$ &  $\beta$ \\
 \hline
$W$ - walking & $0.4011$   & $0.9409$   & $0.0707$ & $0.9162$  \\
             & $(0.0988)$ & $(0.0176)$ & $(0.3327)$ & $(0.0310)$  \\
\hline             
$Y$ - cycling & $0.4006$  & $0.8349$ &  $0.0066$ & $0.9568$ \\
          & $(0.2093)$ & $(0.0336)$& $(0.0066)$ & $(0.0466)$\\
 \hline
 \end{tabular} 
 \caption{Parameters estimated for the number of $W$ walking journeys, and $Y$ cycling journeys.} \label{ParametersWalkCycle}
\end{table}

\subsection*{C - Modal share and income}

The relationship between income and modal share is computed by fitting $C_i \propto Y_i^{\gamma_C}$, where $C_i$ represents car journeys and $Y_i$ represents the income of a city. A similar model is applied to the relationship between income and active mobility (with $\gamma_A$) and public transport (with $\gamma_B$). Results are shown in Table \ref{ParamsIncome}.

\begin{table}
\centering
 \begin{tabular}{|r |cc|} 
 \hline
 Mode & $\delta$ & $\gamma$  \\
 \hline
$A$ - active &  901.5 & -1.435153\\
              & (776.2)  & (0.08349467) \\
\hline             
$B$ - public transport & 177784.2 & -1.435153  \\
                       & (259500) & 0.08349467 \\
\hline
$C$ - cars & 0.004784346 & 0.4545897 \\
           & 0.001005238 & 0.01769237 \\
 \hline
 \end{tabular} 
 \caption{Impact of income on the modal share of cities.} \label{ParamsIncome}
\end{table}

\bibliographystyle{unsrt}

\end{document}